\RequirePackage{fix-cm}
\documentclass[twocolumn]{svjour3}          
\smartqed  
\usepackage{graphicx}

\usepackage[ngerman,english]{babel} 

\usepackage{color} 
\usepackage{ulem}

\usepackage{tabularx}
\usepackage{booktabs}
\usepackage{multirow}

\usepackage{hyperref} 

\usepackage{float} 
\usepackage{textcomp} 
\usepackage{mathptmx}      

\usepackage[round,authoryear]{natbib}
\begin{document}
\title{
Approximation of the emission coefficient for thermography by the combination of geometrical and spectral information - ThermoHead
}
\author{Sebastian Fiedler \and Roland Clau{\ss} \and Hartmut Clau{\ss} \and Stefan Knoblach
}

\institute{mail@dr-clauss.de / Sebastian.Fiedler@FHWS.de
\newline Dr. Clau{\ss} Bild- und Datentechnik GmbH - CLAUSS,
\newline Turnhallenweg 5a, 08297 Zw{\"o}nitz, Germany
\at Stefan Knoblach \& Sebastian Fiedler
\newline University of Applied Sciences
W{\"u}rzburg-Schweinfurt (FHWS), 
\newline R{\"o}ntgenring 8, 97070 W{\"u}rzburg, Germany
}

\date{Received: date / Accepted: date}
\maketitle
\begin{abstract}
	
Thermography allows for the remote measurement of surface temperatures and is widely used for the identification of energy losses, damage detection or quality control. However, thermal imaging is strongly material dependent and therefore measured and real temperatures can differ significantly. The emission coefficient resides between 0 and 1 and changes mainly with the electrical conductivity of the surface. While non-conductive surfaces (stone, wood, glass etc.) have a high emissivity of ca. 0.9, metallic surfaces reside around 0.2. If e.g. an object with a temperature of 100\,$^\circ$C and $\epsilon$\,=\,0.9 gets recorded via thermography, the recording shows a temperature of 90\,$^\circ$C. The same conditions for a metallic surface on the other hand can result in a measured temperature of only 20\,$^\circ$C, so room temperature.

This article introduces a novel method how to approximate the electrical conductivity of surfaces by the combination of geometric and spectral information. A new thermographic panorama system called "ThermoHead" is used to generate high-resolution and geometrically calibrated thermal images of the surrounding. For the geometric information, a 3D point cloud from the very same point of view will be recorded with an arbitrary commercial available terrestrial 3D laser scanner.
Since both data sets have been recorded successively from the exact same point of view, the congruent combination allows to transfer all geometric informations from the terrestrial 3D laser scanner to the thermographic measurement. The radiometric intensity depends on e.g. the angle of incidence of the thermographic measurement, which is now known and can be used to narrow down the electrical conductivity of the measured object.

\begin{otherlanguage}{ngerman}
	\begin{abstract}\sloppy
		Die Thermografie erm{\"o}glicht die Messung von Oberfl{\"a}chentemperaturen aus der Ferne und ist bei der Auffindung von Energieverlusten, Schadensdetektion und der Qualit{\"a}tskontrolle weit verbreitet. Jedoch ist die thermale Bildgebung stark materialabh{\"a}ngig, weshalb sich gemessene Temperaturen deutlich von tats{\"a}chlichen unterscheiden k{\"o}nnen. Der Emissionskoeffizient kann Werte zwischen 0 und 1 annehmen und {\"a}ndert sich haupts{\"a}chlich mit der elektrischen Leitf{\"a}higeit der Oberfl{\"a}che. W{\"a}hrend nichtleitende Oberfl{\"a}chen (Stein, Holz, Glas etc.) eine hohe Emissivit{\"a}t von ca. 0,9 aufweisen, liegen Metalle im Bereich von 0,2. Wenn z.B. ein Objekt mit einer Temperatur von 100\,$^\circ$C und $\epsilon$\,=\,0,9 mittels Thermografie aufgenommen wird, zeigt die Aufnahme eine Temperatur von 90\,$^\circ$C. Unter den selben Bedingungen kann die Messung einer metallischen Oberfl{\"a}che jedoch eine Temperatur von 20\,$^\circ$C anzeigen, demnach Raumtemperatur.
				
		In diesem Beitrag wird eine neuartige Methode vorgestellt, wie sich die elektrische Leitf{\"a}higkeit von Oberfl{\"a}chen durch die Kombination geometrischer und spektraler Informationen bestimmen l{\"a}sst. Ein neuartiges Thermografiepanoramasystem genannt ''ThermoHead'' wird verwendet, um hochaufgel{\"o}ste und geometrisch kalibrierte Thermografiebilder der Umgebung anfertigen zu k{\"o}nnen. F{\"u}r die geometrischen Informationen wird vom selben Standpunkt aus eine 3D-Punktwolke mit einem beliebigen, kommerziell erh{\"a}ltlichen terrestrischen 3D-Laserscanner aufgenommen.
		Da beide Datens{\"a}tze nacheinander vom exakt selben Blickwinkel aus aufgenommen wurden, erlaubt die deckungsgleiche Kombination, alle geometrischen Informationen der terrestrischen 3D Lasermessung auf die thermografische Messung zu {\"u}bertragen. Die radiometrische Intensit{\"a}t h{\"a}ngt z.B. vom Messwinkel der thermografischen Messung ab, der nun bekannt ist und zur Bestimmung der elektrischen Leitf{\"a}higkeit des gemessenen Objekts verwendet werden kann.
	\end{abstract}
\end{otherlanguage}

\end{abstract}

\selectlanguage{english}
\section{Introduction}
\label{sec:1}

In thermal imaging, the infrared radiation of the surface of an object gets recorded with a special camera. The imaging sensor of these thermographic cameras consist of a so called micro bolometer array, which is able to measure radiation in the wavelength $\lambda$\,=\,7\,$\mu$m\,-\,14\,$\mu$m. Micro bolometers absorb radiation, which causes the device to heat up and changes its electrical resistance. In return, this change of resistance can be quantitated and is equivalent to the amount of detected infrared radiation \citep{Basics_TI}. To finally calculate the surface temperature out of the radiometry, some very important variables have to be taken into account \citep{Factors}, which will be discussed in the following:

\subsection{Measurement distance and ambient conditions}
\begin{figure}[h]
	\includegraphics[width=\linewidth]{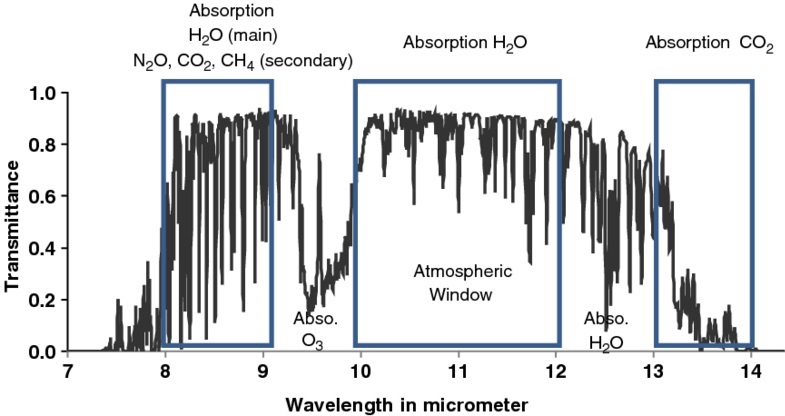}
	\caption{
Atmospheric transmission for thermographic measurements. While the atmospheric window allows for a relatively high transmission for the wavelength $\lambda$\,=\,7\,$\mu$m\,-\,14\,$\mu$m, elevated humidity absorbs infrared radiation and therefore decreases the intensity. Figure taken from Ref.~\cite{H2O}.
	}
	\label{fig:1}
\end{figure}

The amount of detected infrared radiation decreases with increasing distance between thermographic camera and recorded object. The atmosphere is not fully transparent in the wavelengths used for thermography, because polar molecules (CO$\textsubscript{2}$, O$\textsubscript{3}$, CH$\textsubscript{4}$ etc.) can interact with the infrared light and cause absorption, reflection or re-emission. If the ambience shows additionally a high amount of humidity (H$\textsubscript{2}$O is also a polar and therefore infrared-active gas), transmission further decreases [see Fig.~\ref{fig:1} \citep{H2O}]. So for a precise thermographic measurement, the distance between camera and object needs to be considered as well as the humidity. Standard values are 1\,m for the distance between lens and object and an assumed humidity of 50\,\%. The overall influence of distance and humidity can be considered small in close-range scenarios and common ambient circumstances (e.g. no fog).

\subsection{Emission coefficient}
\begin{figure}[h]
	\includegraphics[width=\linewidth]{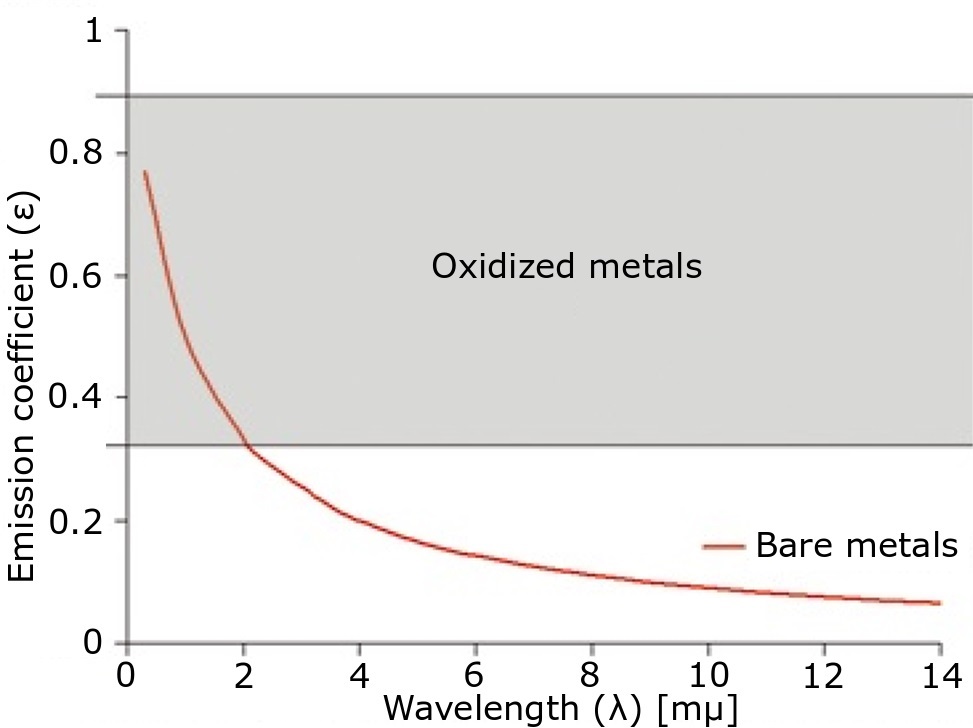}
	\caption{
		Difference in emissivity between electrically conductive and non-conductive surfaces. While oxidized metals and other non-conductors show a high emission over a broad spectrum, metallic surfaces have a low emissivity, which further decreases with increasing wavelength. Figure adapted from Ref.~\cite{Optris}.
		\label{fig:2}
	}
\end{figure}

The emission coefficient is maybe the most critical parameter in thermography. It resides between 0\,$\textless$\,$\epsilon$\,$\textless$\,1 and mainly depends on the electrical conductivity of the measured object. While non-conductive materials (stone, wood, glass etc.) show a high overall emission coefficient of around 0.9, metals can be found at 0.2 or even below [see Fig.~\ref{fig:2} \citep{Optris}]. However, this issue can not be solved by simply switching from long-wave infrared (LWIR: $\lambda$\,=\,7\,$\mu$m\,-\,14\,$\mu$m) to mid-wave infrared (MWIR: $\lambda$\,=\,3\,$\mu$m\,-\,5\,$\mu$m), because the overall radiation intensity decreases along with the wavelength \citep{OPTO}, according to Planck's law of black-body radiation \citep{Planck}.

To calculate the temperature out of the radiometry, the exact emissivity has to be known. Commonly, the emission is set to 1 for all measurements. If an object has been recorded and shows a surface temperature of 90\,$^\circ$C but with $\epsilon$\,=\,0.9, this value has to be multiplied by a suitable factor (here 1.11) to compensate the error from the emissivity.

With a low emissivity comes a high reflectivity [$\epsilon$\,+\,$\rho$\,=\,1 \citep{Optris}]. Not only are hot metallic surfaces often assumed on room-temperature in thermography, in addition, thermal reflections, caused by secondary heat sources, are frequently misinterpreted as the surface temperature itself.
The emission coefficient has the highest impact on thermal imaging and mainly depends on the electrical conductivity of the measured object.

\subsection{Measurement angle}

\begin{figure}[h]
	\includegraphics[width=\linewidth]{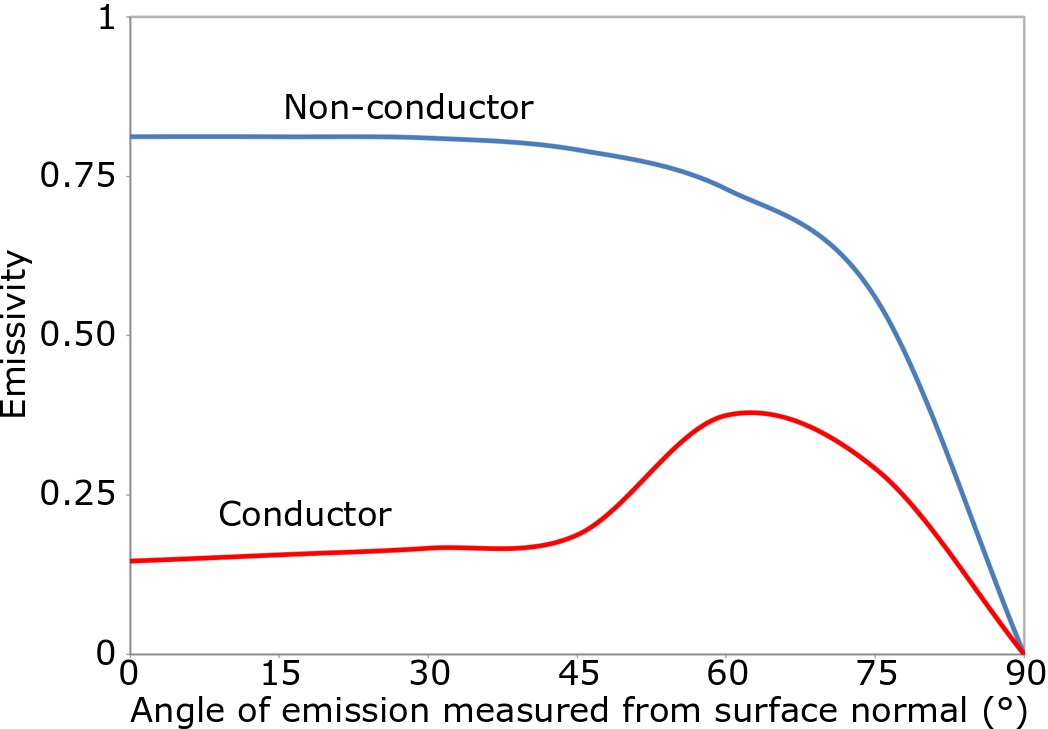}
	\caption{
Not only the emission coefficient differs significantly between conductors and non-conductors, also the emission angle is different. While non-conducting surfaces (electrical insulators / blue) show a high emissivity which decreases at high emission angles, conductors (metals / red) show a low intensity that is increasing at high emission angles. Figure adapted from Ref.~\cite{Neutrium}.
		\label{fig:3}
	}
\end{figure}

The measurement angle is the angle, under which the object has been observed. For most thermographic applications, a hand-held thermography camera is used and the object will be recorded perpendicular, so at an emission angle of 0\,$^\circ$. This is recommended because the emissivity depends on the emission angle, too. While for non-conductive surfaces the emission decreases with higher emission angle, metallic surfaces show a low intensity which is increasing at emission angles between ca. 45\,$^\circ$ and 60\,$^\circ$ [see Fig.~\ref{fig:3} \cite{Neutrium}]. This unique behaviour is the key for our approach.

\section{State of the art}

\subsection{Hand-held thermographic cameras}

Thermography has a wide range of applications, one is detecting energy losses in building physics \citep{Building}. The basic requirement is a temperature difference, which can be achieved due to cold weather outside and heating the building from the inside. The building can be recorded either from the inside or the outside; if there are energy losses (caused e.g. due to leaking windows) the inside will show cold spots or the outside hot spots, respectively. In most cases, absolute temperatures are not required, identifying leaks and their origins is sufficient. This example is maybe the simplest application of thermal imaging, yet, it is highly effective. Further applications are the identification of energy losses in industrial facilities \citep{EiiF}, damage detection \citep{IZPF} or quality control \citep{IPA}.

Normally, thermographic investigations are performed by hand-held devices. To ensure the highest possible measurement quality, the variables discussed in the previous section are ruled out by standard methods e.g. holding the thermographic camera one meter away from the object and perpendicular, assuming an emission coefficient of 1 and a humidity of 50\,\%. To further increase the accuracy of the measured data, the distance to the measured object, the ambient humidity and temperature as well as the emission coefficient can be corrected by special thermographic software \citep{ResearchIR}. Emission coefficients can be extracted from according tables \citep{Emissionsgrad} or calculated by covering the object with a sticker or spray paint with known emissivity or directly measuring the surface temperature with a contact thermometer \citep{Testo}. Although this method is easy to apply and provides a reasonable data quality, it limits the possible applications of thermography in relation to contactless measuring and distance.

\begin{figure}[h]
	\includegraphics[width=\linewidth]{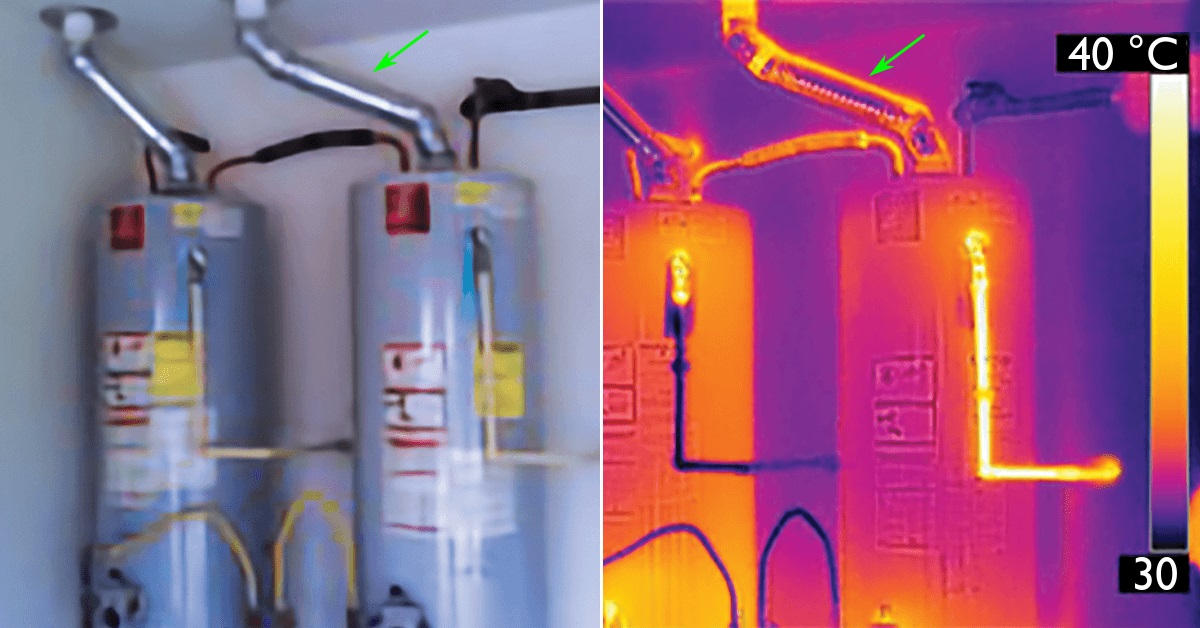}
	\caption{
		High-end hand-held thermography cameras record an RGB-image along with thermography.
		The object can be identified in RGB (e.g. the blank metal pipes on top of the boiler) and according emission coefficients taken out of suitable tables. Please note, the top right metal tube (marked by a green arrow) shows a typical behaviour, increasing intensity with angle. Figure adapted from Ref.~\cite{Premium}.
		\label{fig:4}
	}
\end{figure}

Fig.~\ref{fig:4} shows two images, recorded with the FLIR T1020 \citep{Premium}. This thermographic camera already belongs to the high-end hand-held devices with 3.1\,megapixels (MP) thermal resolution and an additionally built-in RGB camera. The RGB-image allows for a better identification of the thermal image, since the surface material can be estimated: For e.g. the right top pipe (marked by a green arrow), a metallic surface can be assumed (silver shining). In fact, the thermal image shows the typical behaviour with increasing intensity at higher angles (see Fig.~\ref{fig:3}). The problem is, it is assumed that the marked object is a pipe and therefore round, because of the presupposition that pipes go along well with boilers. The measurement angle however as well as the distance to the object can not be extracted out of this image.

\subsection{3D-thermography devices}

Terrestrial 3D laser scanners record their surrounding by emitting laser beams and recording their reflections. The whole device rotates horizontally and has, in most cases, a rotating mirror, which deflects the laser beam along the vertical axis. Both motors have angle decoders, so the position of every emitted laser beam (horizontally and vertically, which is e.g. X and Y) is exactly known. The distance to the object (e.g. Z) will be calculated out of the duration the emitted laser beam needs to return to the device. This will result in a so called 3D point cloud, which is a three dimensional and digital imprint of the surrounding. Additionally, the intensity of the reflected laser beam is recorded. Most of these devices have a built-in RGB camera which colourizes the 3D point cloud of the recorded object in the visible spectrum, too \citep{FARO}. So the resulting data set obtains the 3D geometry of the object (X, Y, Z), the intensity of the reflected laser beam (I) as well as the colour (R, G, B).

\begin{figure}[h]
	\includegraphics[width=\linewidth]{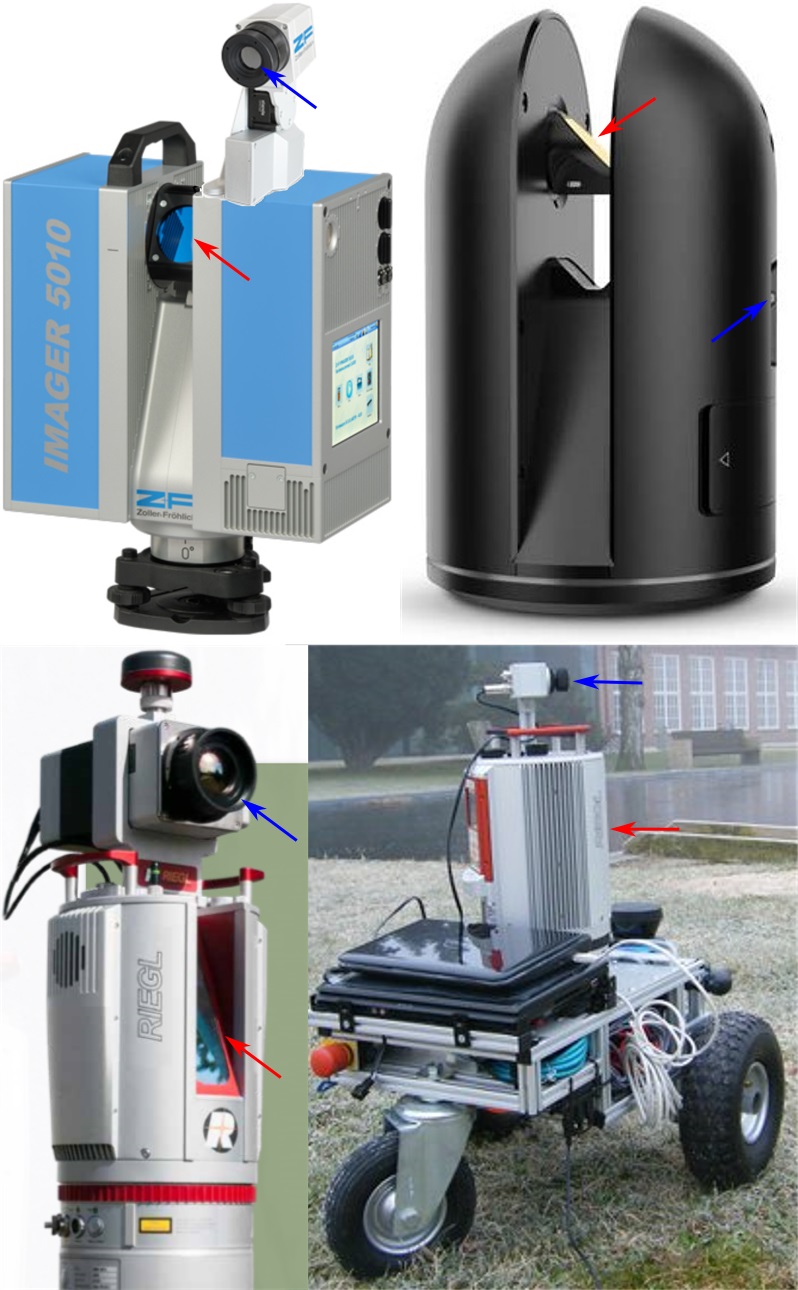}
	\caption{
	The different 3D-thermographic systems: Z+F IMAGER + T-CAM, Leica BLK360, RIEGL VZ-400i and the mobile device ThermalMapper. These systems consist of a terrestrial 3D laser scanner and a thermal camera. The point of view of the laser scanner (red arrow) and the thermal lens (blue arrow) are different, which results in a lack of thermal information at 3D objects (white areas in Fig.~\ref{fig:6}).
	}
	\label{fig:5}
\end{figure}

There are some devices available, which record the 3D geometry along with the thermal information of the environment (see Fig.~\ref{fig:5}). These devices are a combination of a terrestrial 3D laser scanner and a thermographic camera. Since the Leica BLK360 \citep{BLK} and the RIEGL VZ-400i \citep{Riegl} do not record full thermographic panoramas and the ThermalMapper \citep{Borrmann} is not commercially available, the further comparison will be done with the Zoller + Fr{\"o}lich IMAGER + T-Cam \citep{ZF}.

\begin{figure}[h]
	\includegraphics[width=\linewidth]{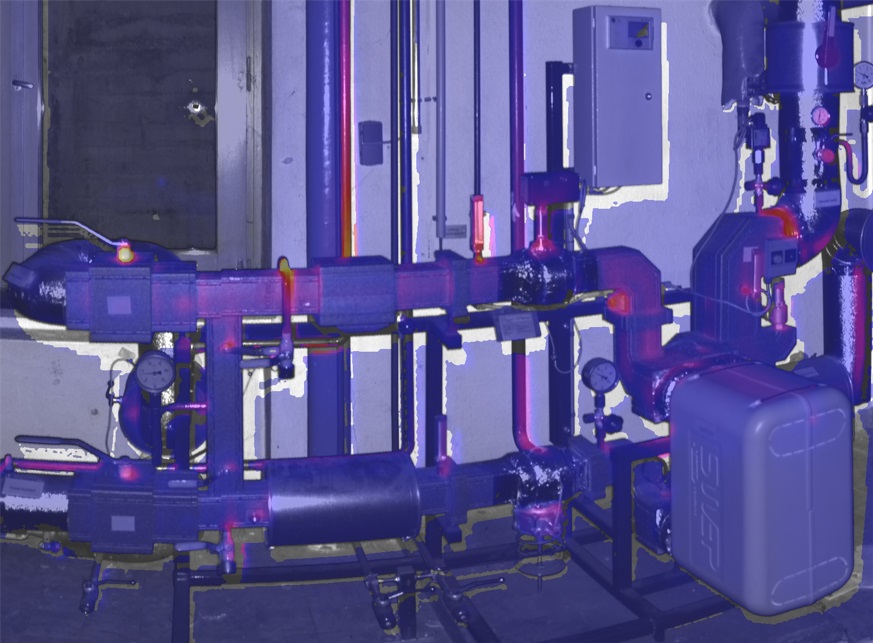}
	\caption{
The 3D-thermographic scan with the Z+F IMAGER 5010 + T-Cam. Due to the different point of view between laser scanner and thermal camera, 3D objects can not be recorded completely. Since the camera is mounted on top left of the device (regarding the scan direction, see Fig.~\ref{fig:5}), white areas appear on bottom right.
	}
	\label{fig:6}
\end{figure}

The IMAGER + T-Cam has been tested extensively \citep{Fiedler} and is the most potent example of the commercially available terrestrial 3D laser scanners with additional thermographic imaging. It records full 360\,$^\circ$\,$\times$180\,$^\circ$ 3D point clouds and thermographic images.
The Z+F IMAGER with T-Cam is a fully automatically working device, which is easy to use and also comes along with an easy data treatment. However, the thermographic panorama has only 3\,MP and can not record 3D objects completely (see Fig.~\ref{fig:6}). The origin resides in the different point of view between terrestrial 3D laser scanner and thermographic camera (marked red and blue in Fig.~\ref{fig:5}).

All these systems (combined 3D-thermographic or hand-held devices) have one common ground: the point of view from laser measurement and RGB-imaging differs from thermography (see different cut-out of the images in Fig.~\ref{fig:4} and white areas in Fig.~\ref{fig:6}). For this reason, angle information of the laser measurement can not be transferred to the thermography unconditionally.

\section{New approach}
For the new approach, 3D point clouds and thermographic images will be recorded successively from the very same point of view \citep{Fiedler}. Therefore, an arbitrary commercially available terrestrial 3D laser scanner can be used to record a 3D point cloud along with the laser reflection intensity and the colour. For this article, measurements have been performed with the FARO FOCUS 3D S120 \citep{FARO}.

High resolution and geometrically calibrated thermographic panorama images in full radiometry have been recorded with the ThermoHead 25, whose height can be matched to every available terrestrial 3D laser scanner with a special adapter.  The centre of the 3D point cloud (X\,=\,Y\,=\,Z\,=\,0) is the terrestrial 3D laser scanner itself, and due to the identical point of view, the centre of the image of the thermographic camera.

\begin{figure}[h]
	\includegraphics[width=\linewidth]{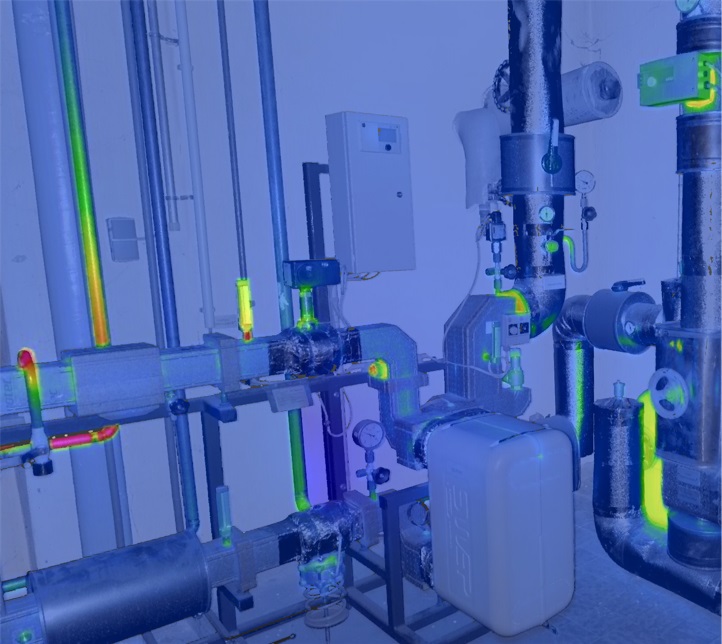}
	\caption{
		Merged 3D point cloud and thermal image, recorded from the same point of view. Not only 3D objects can be measured completely without missing areas (in comparison with Fig.~\ref{fig:6}), bigger cameras can be used which allow a higher geometric and thermal resolution. 
	}
	\label{fig:7}
\end{figure}

\subsection{ThermoHead}

After recording the 3D point cloud, the terrestrial 3D laser scanner gets removed from the stand and the ThermoHead gets attached. Using a suitable height adapter, both devices observe from the very same point of view, because both record around the nodal point. Therefore, both data sets are congruent and can be merged without gaps (compare Fig.~\ref{fig:6} and Fig.~\ref{fig:7}).

\begin{figure}[h]
	\includegraphics[width=\linewidth]{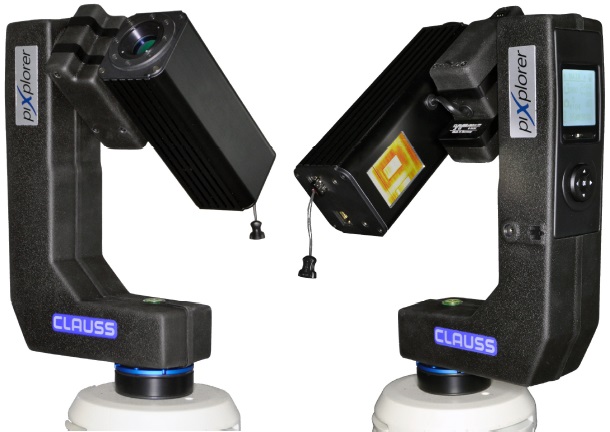}
	\caption{
	The ThermoHead 25 records fully automatically 360\,$^\circ$\,$\times$180\,$^\circ$ thermographic panoramas with a resolution of 25\,MP.
	Due to the geometric calibration, the panoramas can be combined with geometrically stable data sets recorded by e.g. terrestrial 3D laser scanning.
	The complete radiometry is maintained, allowing the data to be further processed in special thermographic software. 
	}
	\label{fig:8}
\end{figure}

\begin{figure*}[h]
	\includegraphics[width=\textwidth]{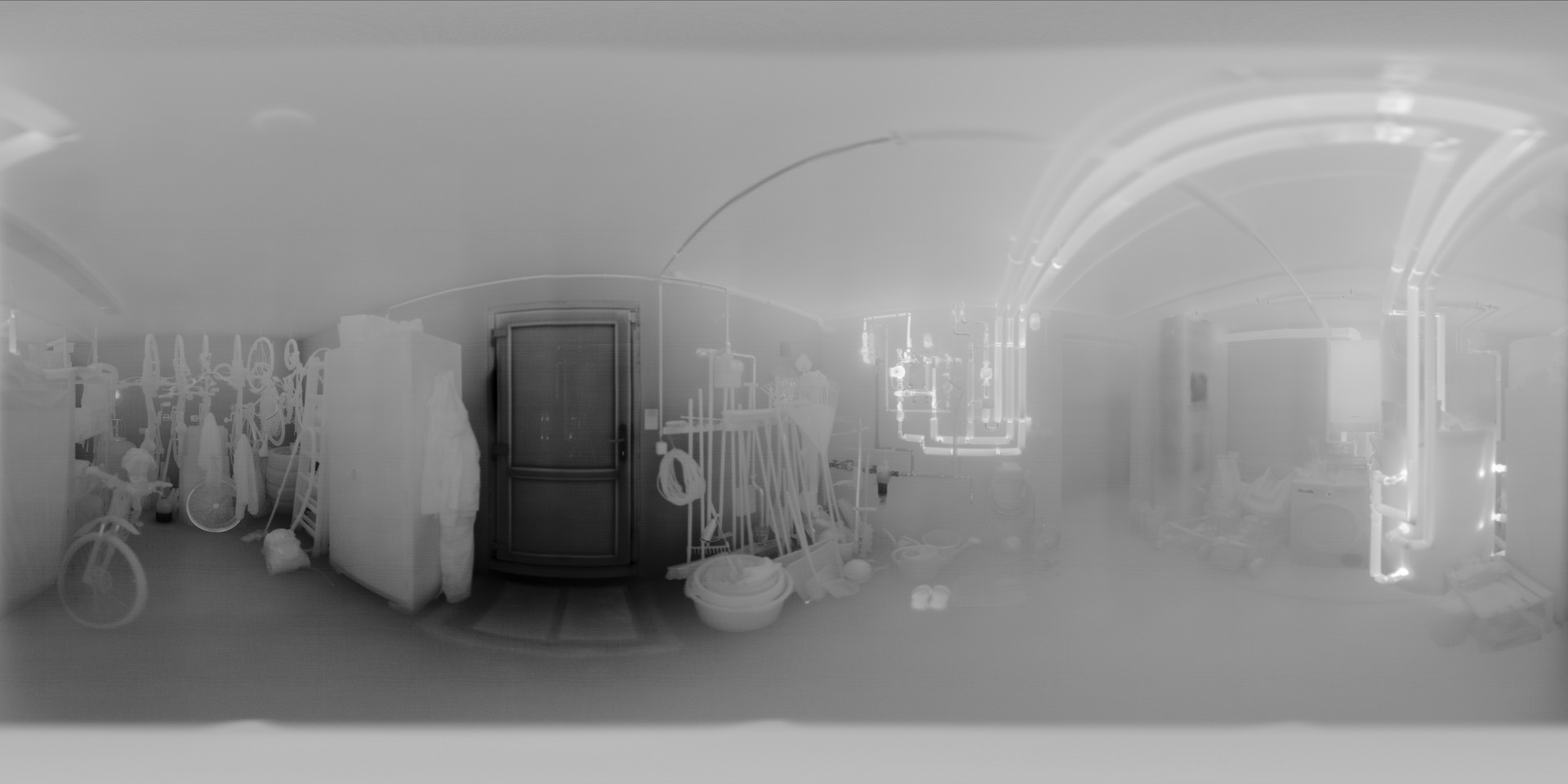}
	\caption{
		Geometrically calibrated thermal panorama image recorded with the ThermoHead 25. The image shows the heating basement of the manufacturer as "Greyscale" from bright (hot) to black (cold) \citep{ResearchIR}. This image can be overlayed with a 3D point cloud or RGB information for a clearer visualisation and better data interpretation.
	}
	\label{fig:9}
\end{figure*}

For the ThermoHead 25 (Fig.~\ref{fig:8}), every camera gets geometrically calibrated by a series of more than 1,000 images \citep{Fiedler}. Due to this calibration, the thermal panorama pictures can be merged with geometrically stable data sets taken from e.g. terrestrial 3D laser scanners. The device fully automatically records 360\,$^\circ$\,$\times$180\,$^\circ$ thermographic panoramas of 25\,MP within 5 minutes. 106 images are taken which can be averaged to increase the thermal resolution. The system can record focussed images from 4,5\,m forward; this working distance can be adjusted to suit particular demands. The complete radiometric information is maintained in a 16\,bit TIF file, which can be further processed in any suitable thermographic software e.g. FLIR ResearchIR \cite{ResearchIR}. A 25\,MP thermal image panorama, recorded with the ThermoHead 25, can be seen in Fig.~\ref{fig:9}.

\subsection{Combination}

Some terrestrial 3D laser scanners do not record RGB information. Therefore, certain software implemented algorithms to merge 3D point clouds with high-resolution RGB panoramas, recorded by panorama devices. The approach to combine 3D point clouds with thermographic panoramas is comparable.
Good results have been archived with Trimble RealWorks \citep{Trimble}. The quality of the geometric calibration of the thermal panorama image is ensured by checking the fitting of geometry and thermography in the whole 3D thermographic data set in near- and far-field. This approach can be found in detail in Ref.~\cite{Fiedler}. For the further approach, the automatic colourizing option "RealColor" in Trimble RealWorks has been used.

\section{Experiment}

\subsection{Setup}

\begin{figure}[h]
	\includegraphics[width=\linewidth]{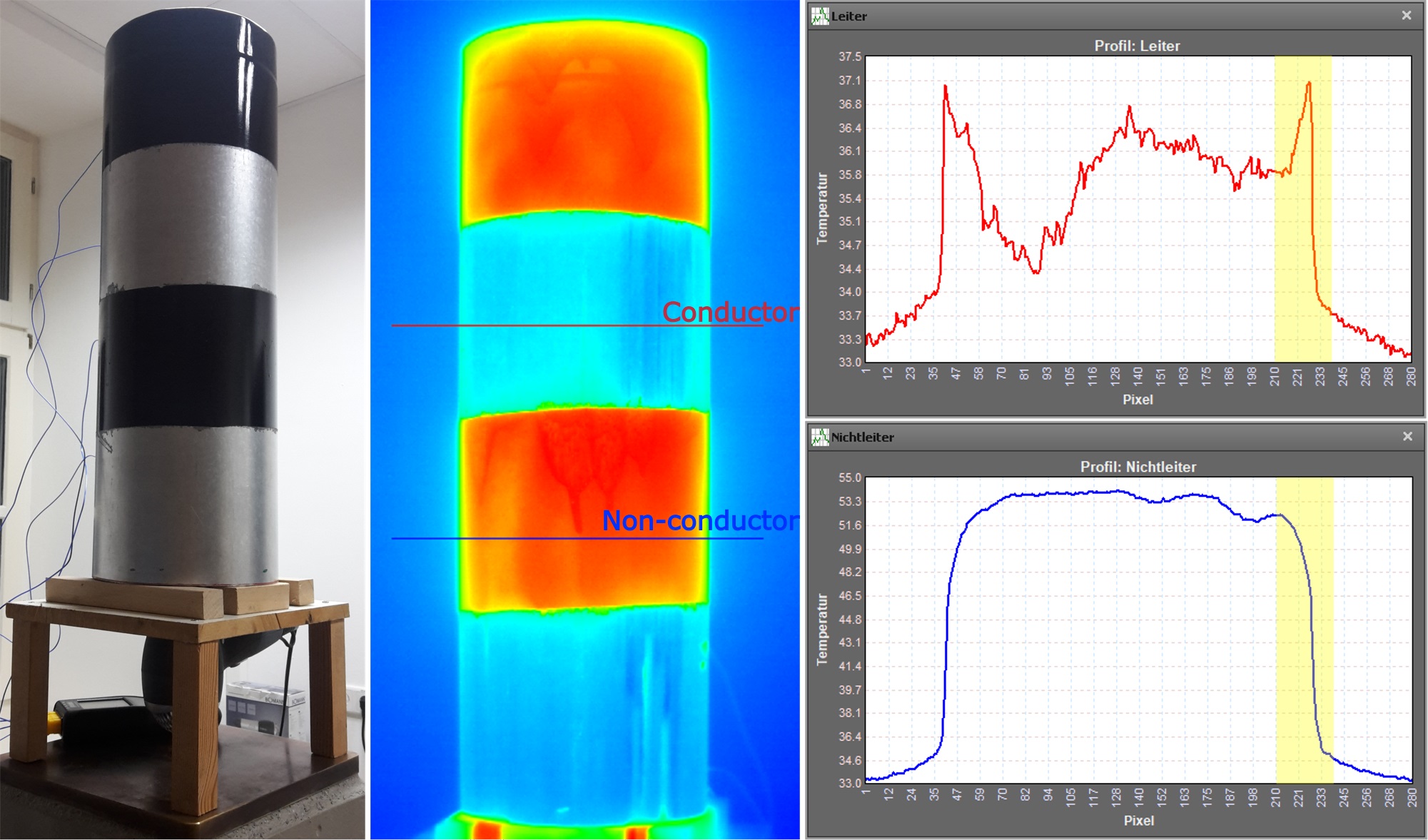}
	\caption{
	The test object is a metal cylinder that can be heated to ca. 57\,$^\circ$C from the inside and shows surfaces with two different emission coefficients.
	In thermography, the conductive surface (silver surface / red line) shows temperatures of around 37\,$^\circ$C while the non-conductive surface (black surface / blue line) shows temperatures of ca. 54\,$^\circ$C. The typical behaviour at angles between 45\,$^\circ$ and 60\,$^\circ$ (increasing intensity for conductors, decreasing for non-conductors, compare with Fig.~\ref{fig:3}) has been marked yellow.
	}
	\label{fig:10}
\end{figure}

For the experimental part, a test object has been manufactured (see Fig.~\ref{fig:10}). Therefore, a metal cylinder with a diameter of 20\,cm has been used, which is closed on top. The cylinder is on a wooden stand, which allows heating it from the inside via hot air blower. Half of the cylinder has been covered by black spray paint (non-conductor with a high emission coefficient) while the other half remains uncovered, resulting in 2\,$\times$\,2 areas. One K-type thermocouple \citep{TC41} has been attached to every area to record the temperature and ensure a homogenous heating of the test object. Temperatures have been found at 57.3($\pm$ 0.3)\,$^\circ$C. Thermography shows a temperature of around 54\,$^\circ$C for the areas covered with black paint and 37\,$^\circ$C for the metallic areas.
Calculating the emission coefficient for both differently covered parts, the black paint shows an $\epsilon$\,$\approx$\,0.9, which is reasonable. The same method for the metallic part however would lead to an emission coefficient of $\epsilon$\,$\approx$\,0.6, which is surprisingly high for a metal, but the impact of ambient reflections have not been taken into account.
A terrestrial 3D laser scan as well as a thermal panorama image has been recorded from the test object at a distance of 2\,m.

When analysing the temperature distribution along one dimension (line profile), the typical behaviour, low but increasing temperature for conductors (red line) and high but decreasing temperatures for non-conductors (blue line), which has been shown in Fig.~\ref{fig:2}, can be observed and has been marked yellow for the according angles (Fig.~\ref{fig:10}). With this information, the different surface materials can be identified as conductive and non-conductive. Both line profiles show an uneven course: the origin for the non-conductor is a inhomogeneous thickness of the spray paint while for the conductor secondary heat sources are present. However, for the thermal panorama system, the whole surrounding of the object gets recorded, too, allowing for the identifications of further heat sources \citep{Fiedler}.

\begin{figure}[h]
	\includegraphics[width=\linewidth]{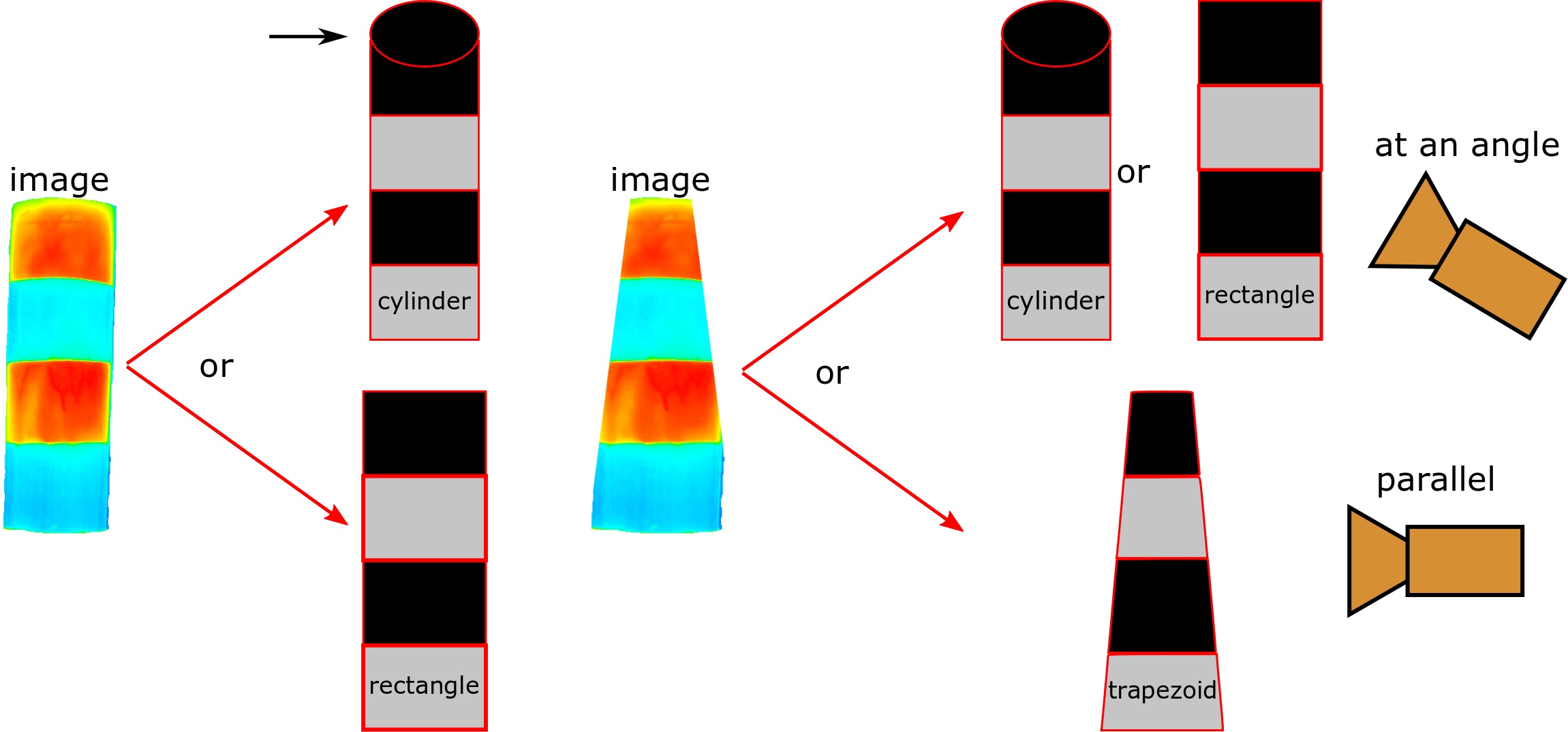}
	\caption{
		When taking a thermal image of a 3D object, all information concerning distance to the object, geometry and observing angle get lost due to the 2D-projection. The thermal image (Fig.~\ref{fig:10}) could have a 3D (cylinder) or a 2D (rectangle) origin. If the object is not recorded perpendicular, even more initial situations are possible.
	}
	\label{fig:11}
\end{figure}

But, as mentioned before, this is only possible in the current situation because the prior knowledge that the object is round has been included. Again, the measurement angle as well as the distance to the object can not be extracted out of this thermographic image. Since the true geometry of the object is unknown, further initial situations are possible, e.g. the object was not 3D (cylinder) but 2D (square) or if it appears as a trapezoid it could be a trapezoid recorded parallel or a parallel object recorded at an angle (see Fig.~\ref{fig:11}). To determine the emissivity, the recorded thermographic intensity has to be analysed in dependency of the emission angle (Fig.~\ref{fig:3}). This is currently not possible, because an image is a 2D-projection of a 3D object, therefore, the necessary information are not maintained.

\subsection{Result}

\begin{figure}[h]
	\includegraphics[width=\linewidth]{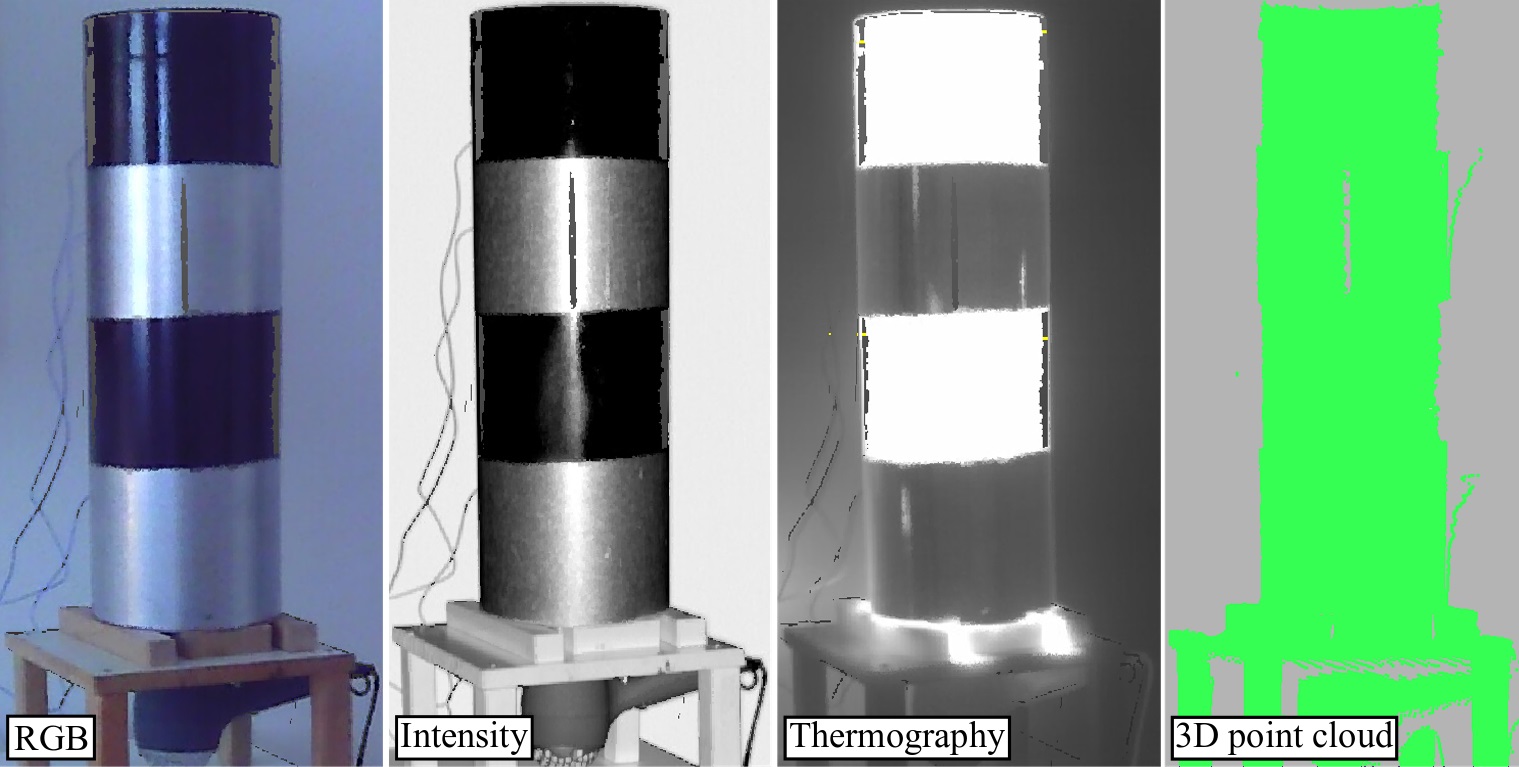}
	\caption{
The test object shown in RGB, intensity, thermography and the 3D point cloud. The information about the geometry of the object, which has been lost due to the 2D-projection in thermal imaging, has been restored by the terrestrial 3D laser measurement.
	}
	\label{fig:12}
\end{figure}

Measurements have been combined according to the previous section. The accuracy of fit has been verified by several thermographic markers distributed in the room around the test object. The result is a 3D data set including the distance to and geometry of the object (X, Y, Z), laser reflection intensity (I), the colour (R, G, B) and the thermographic information (T) \citep{3DT}, \citep{eDIan}, shown in Fig.~\ref{fig:12}.

\begin{figure}[h]
	\includegraphics[width=\linewidth]{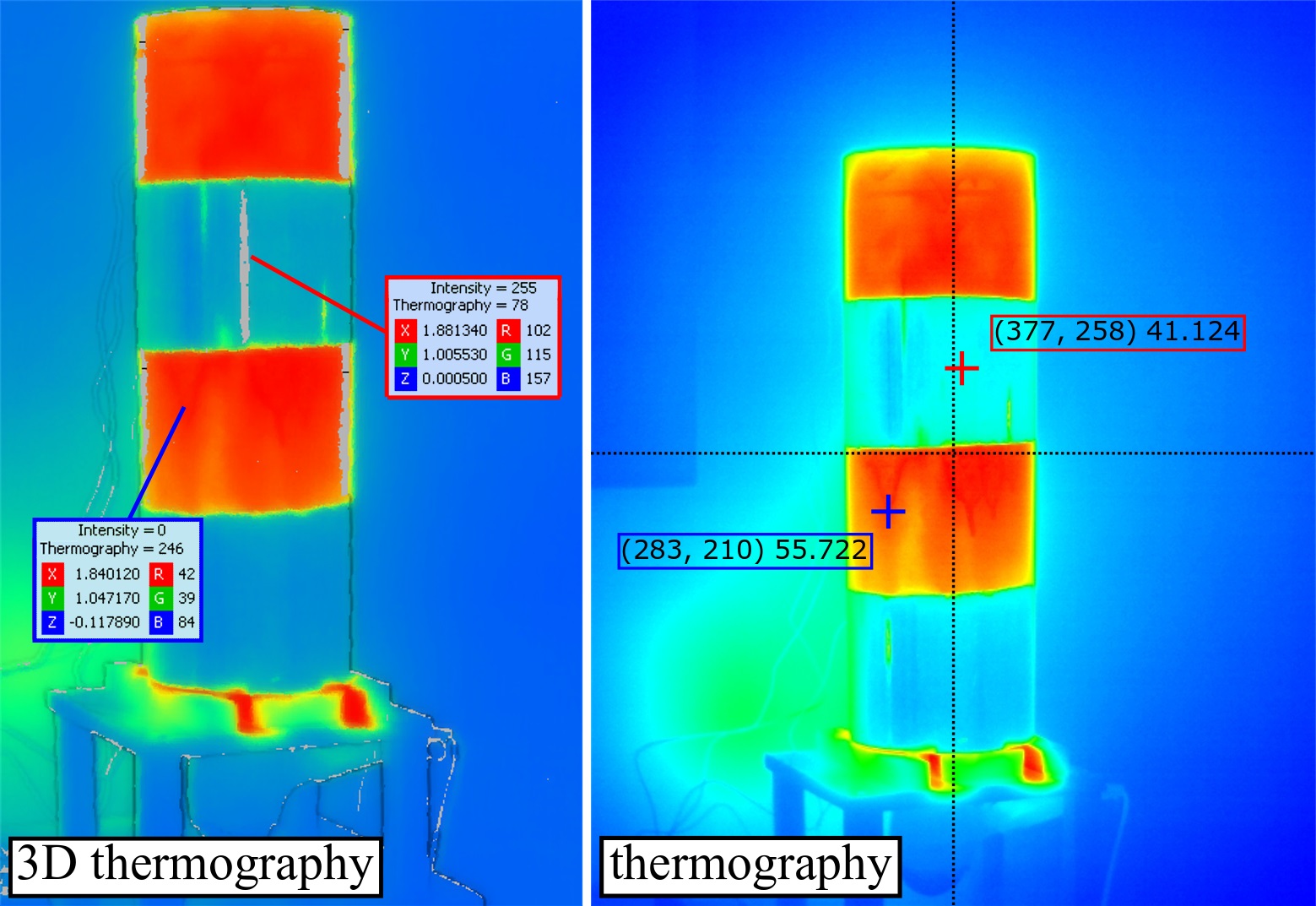}
	\caption{
In 3D thermography, for every pixel the distance to the laser scanner can be extracted as well as the RGB value, the laser reflection intensity and the thermography. Vice versa, for every pixel of the thermographic image it is possible to determine the distance to the measured object as well as the original radiometric values.
	}
	\label{fig:13}
\end{figure}

Since terrestrial 3D laser measurement and thermography have been recorded successively from the exact same point of view, the distance between thermal lens and object can be extracted for every pixel directly out of the 3D point cloud. This is shown in the left hand side of Fig.~\ref{fig:13}, where for two different positions (conductor marked red and non-conductor marked blue) the values for RGB, intensity and thermography have been extracted. For the current example, thermography has been reduced to 256 values (8\,bit), so it can be stored and displayed in CloudCompare \citep{CloudCompare}.
Vice versa, when analysing the thermographic images in FLIR ResearchIR MAX (Fig.~\ref{fig:13} right), the full radiometric values are accessible for every pixel. Here, the coordinate of the corresponding pixel in the thermographic image is shown as two comma separated values in brackets (size of the shown cut out is 640\,$\times$\,512 pixels) and the radiometric value is displayed as a number with five digits (16\,bit correspond to 65.536 increments).

The values from Fig.~\ref{fig:13} have been extracted and put in Tab.~\ref{tab:1}. Please note that the black surface (low RGB but high thermographic values) also shows a low laser reflection intensity. The intensity is also an indicator for the surface roughness in the near visible spectrum ($\lambda$$\textsubscript{FARO S120}$\,=\,905\,nm) but not in the used infrared wavelength.

\begin{table*}[h]
	\caption{Information extracted from Fig.~\ref{fig:13} for the metallic (red) and the insulating (blue) surface. Every point includes the 3D-position (X,Y,Z), the colour (R,G,B), laser reflection intensity (I) and the thermography (T; reduced to 256 values). Furthermore, the corresponding pixels in the thermographic image (X,Y) can be identified and the original radiometric values (Rad) extracted.}
	\begin{tabular*}{\textwidth}{@{\extracolsep{\fill}}*9l@{}}
	\textbf{Surface} & \textbf{X ~~~~~~~~~~~~Y ~~~~~~~~~~~~~Z} (3D)      & \textbf{R} & \textbf{G} & \textbf{B} & \textbf{I} & \textbf{T} & \textbf{X ~~~~~Y} (2D) & \textbf{Rad} \\
	Metallic         & 1.88134 ~~1.00553  ~~~0.00050                     & 102        & 115        & 157        & 255        & 78         & 377 ~~258              & 41.124               \\
	Insulator        & 1.84012 ~~1.04717 ~~-0.11789                      & 42         & 39         & 84         & 0          & 246        & 283 ~~210              & 55.722               \\
	\end{tabular*}
    \label{tab:1}
\end{table*}

\begin{figure}[h]
	\includegraphics[width=\linewidth]{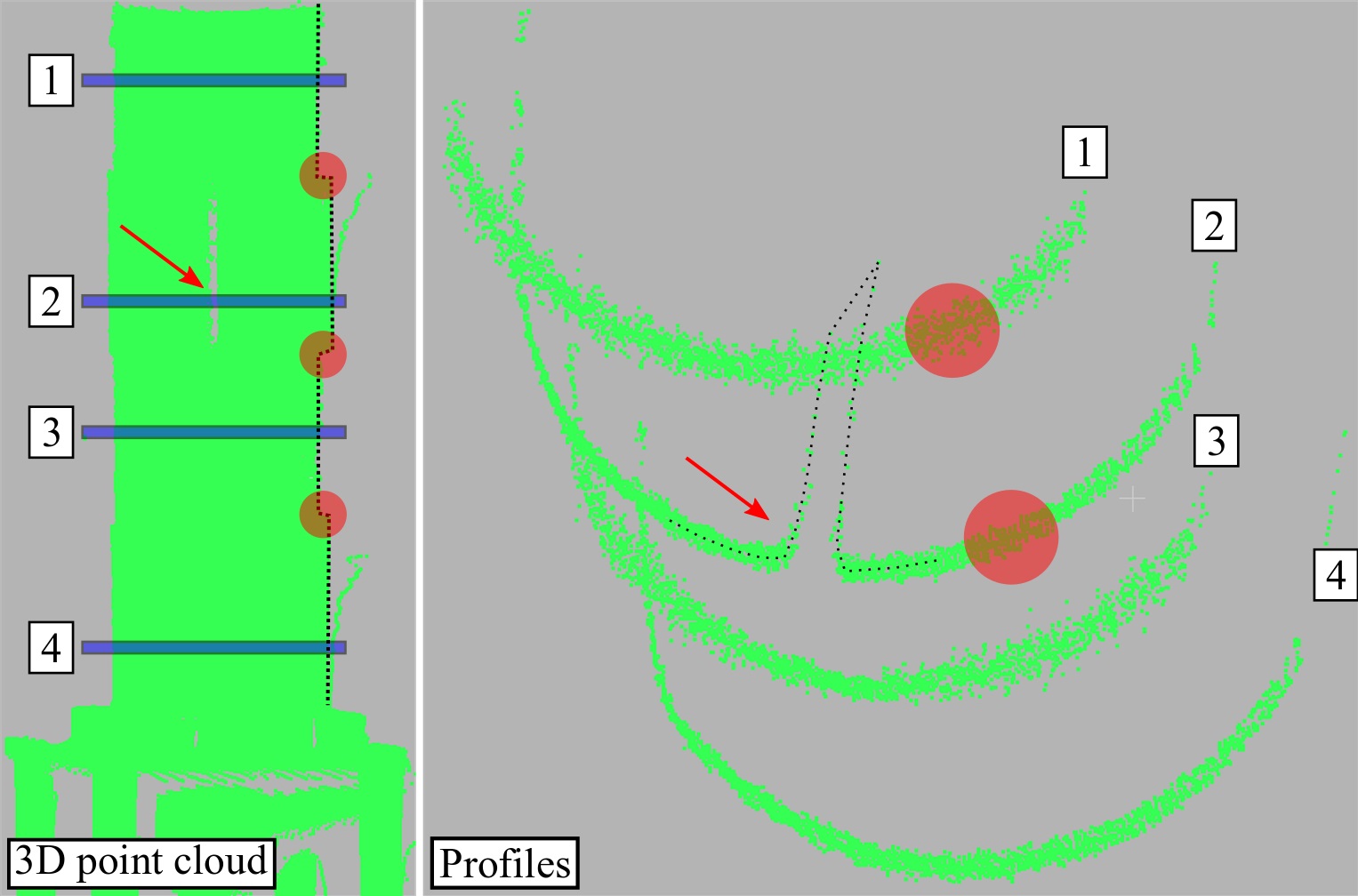}
	\caption{
When analysing the 3D point cloud, both surfaces show different characteristics: The black surface (slice 1 and 3) show a higher noise and a less dense point cloud compared to the metallic surface (slice 2 and 4) marked by red circles. The reason is the lower intensity of reflected laser beams. Additionally, total reflections occur (marked by a red arrow) when measuring perpendicularly on surfaces with high reflection.
	}
	\label{fig:14}
\end{figure}

When taking a closer look at the 3D point cloud (Fig.~\ref{fig:13}), different characteristics can be observed for the metallic and painted surface. At high angles, the object seems to be smaller for the black surface (marked red on the left hand side of Fig.~\ref{fig:14}). The reason is the lower laser reflection intensity. Less laser beams are reflected and therefore recorded by the terrestrial 3D laser scanner, which also leads to a difference in the overall quality of the point cloud, where the non-conductive area shows a higher noise (marked red on the right hand side of Fig.~\ref{fig:14}).

Furthermore, highly reflecting surfaces show a very unique behaviour when recorded perpendicularly: When following the intensity of the metallic surface, a slit can be seen in the 3D point cloud (marked by a red arrow). This happens when a total reflection appears, where the points get wrongfully put inside of the object. So two very unique characteristics can be differentiated here; a less dense point cloud with higher noise for the black surface and low noise with a total reflection for perpendicular measurements for the metallic one. Also, this intensity is depended on the measurement angle. This additional information can be used to further narrow down the electrical conductivity of the measured surface.

To finally reconstruct the geometry of the object and determine the measurement angles, tree different approaches have been proven purposeful so far:
Positions can be extracted directly out of the 3D point cloud (Fig.~\ref{fig:14}) which describe the curvature of the object. By comparing the coordinates of the selected points, the course of the object relative to the thermographic lens can be determined. This method is less precise but quick.
A 3D point cloud consists of single points, which are not related to each other. When connecting the neighbouring points to each other (meshing), a polygonal model can be calculated. Out of this 3D model, all relevant geometric properties can be extracted. This approach has proven time consuming, yet the most precise method.

\begin{figure}[h]
	\includegraphics[width=\linewidth]{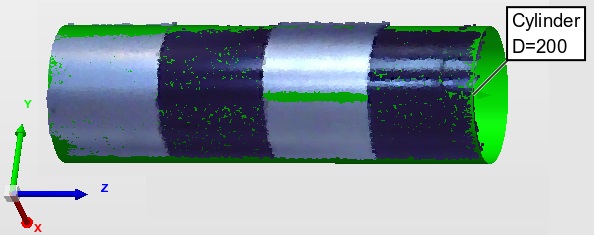}
	\caption{
Reconstruction in PolyWorks. The 3D point cloud of the object can be e.g. meshed (polygonalized) to extract the curvature. Here, the object has been completed by a standard geometry which is a cylinder with 20\,cm diameter. Since the centre of the point cloud is the terrestrial 3D laser scanner and therefore the lens of the thermal camera, all angles can now be extracted.
	}
	\label{fig:15}
\end{figure}

For the further approach, the 3D point cloud has been imported in PolyWorks \citep{PolyWorks} for the reconstruction of a 3D model (Fig.~\ref{fig:15}). Therefore, all points of interest have been selected and processed to a standard geometry (cylinder). Since the centre of the 3D point cloud is the terrestrial 3D laser scanner itself and due to the identical point of view the thermographic camera, all distances and angles can be extracted.

Now extracting the intensity of the thermographic measurement as a line profile (Fig.~\ref{fig:10}) and extracting the curvature of the object (Fig.~\ref{fig:14} or Fig.~\ref{fig:15}) allows for the identification of the electrical conductivity (Fig.~\ref{fig:3}).

\begin{figure*}[h]
	\includegraphics[width=\textwidth]{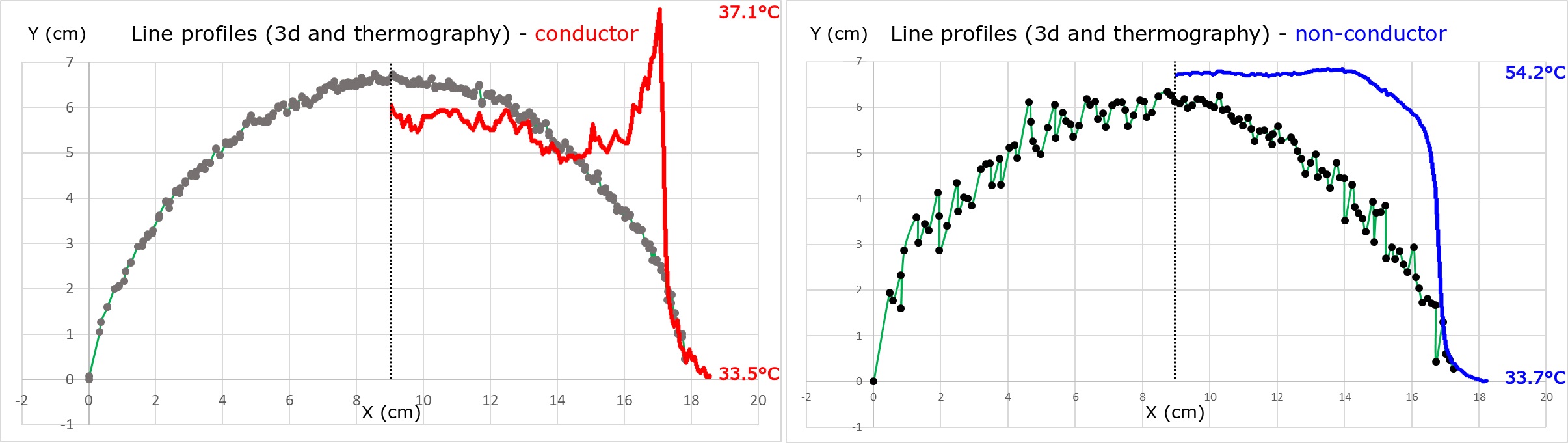}
	\caption{
Overlay of the line profiles of the 3D point cloud (Fig.~\ref{fig:14} slice 3 and 4) and the corresponding line profiles of the thermography image (analogue to Fig.~\ref{fig:10}). X and Y axis in centimetres. The dashed line indicates the 0\,$^\circ$ measurement angle. For the electrical conductive surface (left) the typical increase in intensity can be matched to high recording angles. For the non-conductive surface (right) the high angles show the typical decrease in recorded thermographic intensity. Please note the diameter is smaller than the original object due to limitations of the scan. Also the black painted surface of the non-conductor (right) shows the higher noise, as has been shown in Fig.~\ref{fig:14}.
	}
	\label{fig:16}
\end{figure*}

By the direct comparison of the line profiles of the 3D point cloud (Fig.~\ref{fig:16}) and the thermography, the electrical conductivity of the surface can be determined. For this example, the areas 3 and 4 in Fig.~\ref{fig:14} has been chosen. Now the increasing thermographic intensity can be localized to the curvature of the object and therefore higher emission angles, which identifies the surface as an electrical conductor (see Fig.~\ref{fig:3}). Please note that the 3D point cloud of the 20\,cm test object only shows a diameter of 18\,cm, since a measurement of 90\,$^\circ$ relative to the instruments is not possible.
The diameter of the non-conductive surface is even smaller, in accordance to the observation that the black colour reflects less laser rays and shows a greater overall noise.
The temperatures shown in Fig.~\ref{fig:3} are not the real surface temperatures, but the ones the thermographic software calculates out of the radiometric values.

\section{Summary and conclusion}

In this manuscript, an new approach to approximate the emission coefficient for thermal imaging has been presented. Therefore, 3D geometric informations of an object have been combined with different spectral informations. A novel thermographic panorama system called ThermoHead has been used to generate high-resolution and geometrically calibrated thermographic panorama images from the exact same point of view as the terrestrial 3D laser scan. Both data sets can be combined congruently. The method has been carried out as followed:

\begin{itemize}
	\item Recording 3D point cloud with an arbitrary terrestrial 3D laser scanner.
	\item Recording a high-resolution and geometrically calibrated thermographic panorama image with full radiometric information from the exact same point of view by the ThermoHead.
	\item Combine both data sets congruently; the point of view of the thermographic camera is now identical with the terrestrial 3D laser scanner.
	
	\item Measuring distance between object and thermographic camera can be extracted out of the 3D point cloud.
	\item Measuring angle of the object can be extracted out of the reconstruction of the object, e.g meshing the point cloud or including a standard geometry.
	\item The electrical conductivity of the surface can be determined by comparing the intensity of the thermographic image (line profile) with the curvature and alignment of the object.
	\item The emission coefficient can be further narrowed down by analysing the RGB and laser reflection intensity values for the thermographic pixel.
\end{itemize}

The combination of both data sets allows to identify the emission coefficient more precise. The distance to the object, its laser reflection intensity and the RGB information can be determined for every thermographic pixel. Due to the reconstruction of the object, the measurement angle is available. By comparing a line profile of the thermographic measurement with the curvature of the object at the same position, the electrical conductivity of the surface can be distinguished. In this manuscript, the electrical conductivity of the surface, which has the highest impact on the emission coefficient, could be distinguished between conductor and non-conductor. The new approach is also suitable to determine the electrical conductivity of surfaces contactless in e.g. far distances. Furthermore, since the geometry of the reflections are available, too, ray-tracing becomes possible.
Currently the FHWS and CLAUSS are working to increase the geometric resolution from the ThermoHead 25 to 100\,megapixels.

\begin{acknowledgements}
We thank the German Federal Ministry of Education and Research (BMBF grant number 03FH021PX5) and the German Federal Ministry of Economic Affairs and Energy (BMWi grant number ZF4662502JA9) for their support.
Thanks to all employees from the mechanical engineering department, surveying and the geo group.

\end{acknowledgements}

\bigskip
\bf{Accessed 21.09.2020}

\end{document}